\documentclass[pra,preprint,showpacs,superscriptaddress]{revtex4}
\usepackage{graphicx}

\begin{document}

\title{Non-Markovian Decay of a Three Level Cascade Atom in a Structured
Reservoir }
\author{B.J.~Dalton}
\affiliation{Centre for Atom Optics and Ultrafast Spectroscopy, Swinburne University,\\
Hawthorn, Victoria 3122, Australia}
\affiliation{Department of Physics and Astronomy\\
University of Sussex, Falmer, Brighton, BN1 9QJ, UK}
\affiliation{Department of Physics, University of Queensland, St Lucia,\\
Queensland 4072, Australia}
\author{B.M.~Garraway}
\affiliation{Department of Physics and Astronomy\\
University of Sussex, Falmer, Brighton, BN1 9QJ, UK}
\date{\today }

\begin{abstract}
The dynamics of a three level atom in a cascade (or ladder) configuration
with both transitions coupled to a single structured reservoir of quantized
electromagnetic field modes is treated using Laplace transform methods
applied to the coupled amplitude equations. In this system two photon
excitation of the reservoir occurs, and both sequences for emitting the two
photons are allowed for. An integral equation is found to govern the complex
amplitudes of interest. It is shown that the dynamics of the atomic system
is completely determined in terms of reservoir structure functions, which
are products of the mode density with the coupling constant squared. This
dependence on reservoir structure functions rather than on the mode density
or coupling constants alone, shows that it may be possible to extend
pseudo-mode theory to treat multiphoton excitation of a structured
reservoir---pseudo-modes being introduced in one-one correspondence with the
poles of reservoir structure functions in the complex frequency plane. A
general numerical method for solving the integral equations based on
discretising frequency space, and applicable to different structured
reservoirs such as high Q cavities and photonic band gap systems, is
presented. An application to a high Q cavity case with identical Lorentzian
reservoir structure functions is made, and the non-Markovian decay of the
excited state shown. A formal solution to the integral equations in terms of
right and left eigenfunctions of a non-Hermitean kernel is also given.

The dynamics of the cascade atom, with the two transitions coupled to two
separate structured reservoirs of quantized electromagnetic field modes, is
treated similarly to the single structured reservoir situation. Again the
dynamics only depends on reservoir structure functions. As only one sequence
of emitting the two photons now occurs, the integral equation for the
amplitudes can be solved analytically. The non-Markovian decay of the
excited state is shown for the same high Q cavity case of identical
Lorentzian reservoir structure functions, and differs from that for the
single reservoir situation.
\end{abstract}

\maketitle

\section{Introduction}

\label{sec:intro}


The quantum optical behaviour of atomic systems coupled to a continuum of
the quantized electromagnetic field modes has been studied since the early
days of quantum physics. The quantum electromagnetic field is a large
system, which can be treated as a bath or reservoir. In most cases the
atom-field coupling constants and the electromagnetic field mode density are
slowly varying functions of frequency, and the dynamics of the atomic system
can be treated via Markovian master equations \cite{masters,Barnett97}, or
equivalent methods such as quantum Langevin equations (see e.g.\ \cite%
{Barnett97} for details of these standard methods). These techniques are
based on quantum electromagnetic field states with no special distinction
for any particular mode in terms of photon occupation number, such as
thermal states or broad-band squeezed states. Naturally, if one mode of the
electromagnetic field was in a special state, such as a large amplitude
coherent state (as in the case where the atom is also coupled to a laser
field), then this special mode and the atomic system would be treated as a
small quantum system with the remaining modes constituting the reservoir, so
that Markovian behaviour would still apply for the small system.

In certain cases however, such as for atoms in high Q cavities or in
photonic band gap materials, either the coupling constants or the mode
density (or both) are no longer slowly varying functions and standard
Markovian master equation methods are no longer valid (see \cite%
{Lambropoulos00} for a recent review). A number of non-Markovian methods
have been formulated, see, for example, references in Ref.\ \cite{Dalton01a}%
. These include non-Markovian master equations \cite%
{Zwanzig64,Nakajima58,Barnett00}, the time-convolutionless projection
operator master equation \cite{Shibata77}, Heisenberg equations of motion 
\cite{Vats98,Cresser00}, stochastic wave-function methods for non-Markovian
processes \cite{Breuer99,Strunz99,Jack99,Molmer99,Quang97,Stenius96},
methods based on the essential states approximation or resolvent operators 
\cite{Lambropoulos00,Paspalakis99,Bay97,Law02}, the pseudo-mode approach 
\cite{Garraway97,Imamoglu94}, Fano diagonalization \cite{Fano61} (and \cite%
{Jeffers00,Dalton01a}), and various short-time approximations \cite%
{Braun01,Privman02}. The last four approaches are easiest to apply,
providing clear physical insight into the processes involved.

One such method is that of pseudo-mode theory \cite{Garraway97,Imamoglu94}.
This method was developed for the case of a two level atom coupled to a
structured electromagnetic field reservoir in the vacuum states and was then
restricted to single photon excitations of the reservoir. The treatment
started from the time dependent state vector for the atom-field system,
written as a linear combination of one photon states with the atom in the
ground state and vacuum states with the atom in the excited state. The basis
of the method was that the atomic dynamics only depended in this case on the
behaviour of a single function, the reservoir structure function, defined as
the product of the mode density and the square of the coupling constant. The
complex frequencies and residues of the poles of this function in the lower
half complex frequency plane enabled so called pseudo-modes to be
introduced, one for each of the finite number of poles. The non-Markovian
equation for the complex amplitude of the state with the atom excited (and
the field in the vacuum state) could then be replaced by Markovian equations
involving the \textit{finite} number of pseudo-mode amplitudes together with
the amplitude for the excited atomic state. The pseudo-modes are originally
related mathematically to the reservoir structure function, but in some
cases their physical origin can be explained. For the case of the atom in a
high Q cavity, where the coupling constants vary rapidly near the cavity
resonance frequencies whilst the mode density is slowly varying, the
pseudo-modes can be interpreted \cite{Dalton01a} in terms of the cavity
quasimodes \cite{Dalton99a}. For the case of an atom in a photonic band gap
system, no pseudo-mode theory is yet available, though a treatment in terms
of quasimodes \cite{Dalton02a} can be used to account for the frequency
dependence of the coupling constants and mode densities. A treatment of
superradiance in a photonic band gap continuum \cite{Bay98b} is based on the
idea of replacing the photonic band gap system by a pair of degenerate
cavity modes coupled to the multi-atom system and with each other, one of
the modes being also coupled to a Markovian bath. In terms of the treatment
in \cite{Dalton01a}, such a case would produce a Fano-profile reservoir
structure function, with the Fano window representing the photonic band gap.
The two cavity modes would correspond to two pseudo-modes. The problem for
photonic band gap situations is that the mode density is actually a
discontinuous function of the frequency, and thus the reservoir structure
function would not have a finite number of simple poles, though approximate
representations of the reservoir structure function in such a form might be
found.

Leaving aside the difficulties associated with the pseudo-mode theory for
photonic band gap systems, it would be desirable to see if pseudo-mode
theory could be extended to cases where multiple photon excitation of the
structured reservoir is involved, as the original treatment \cite{Garraway97}
only covers single photon excitation. The limitation of current treatments
to the single photon excitation case has been noted in Ref.~\cite%
{Lambropoulos00}, but some work has been carried out on cases of multiphoton
excitation of the reservoir. Such a multiphoton situation would apply if the
two level atom was replaced by a three level cascade (or ladder) system,
with an initial condition of the atom in the uppermost state and no photons
present in the electromagnetic field. The two-step decay will generate
electromagnetic field states with two photons present. Another case of
multiphoton excitation occurs for an excited two level atom coupled to a
defect mode containing one photon, the atom also being coupled to a photonic
band gap continuum. Beginning with the essential states approximation, a
numerical method based on replacing the density of modes by a discretised
model has been used in this latter situation \cite%
{Nikolopoulos99,Nikolopoulos01} and in the case of the cascade system \cite%
{Nikolopoulos00}. Similar numerical methods have also been used to treat
stimulated emission in a photonic crystal \cite{Konopka99}. The cascade
system case with one transition coupled near-resonantly to the edge of a
photonic gap (and the other coupled to a flat continuum) has also been
treated via the resolvent operator method in Ref. \cite{Bay98c}. Although
the treatment is analytic, this feature results from being able to ignore
processes in which the two emitted photons are produced in a different
sequence---a reasonable approximation if the two transition frequencies are
very different. However, a more general analytical method would be
desirable, and therefore we aim to see if pseudo-mode theory can be extended
to treat the multiphoton excitation case without having to make assumptions
about the order in which the photons are produced. Whether an extension is
possible involves first showing that the atomic dynamics only depends on the
behavior of reservoir structure functions---in a cascade system we would
expect there to be more than one reservoir structure function, since two
coupling constants are present. A next step would be to then introduce
suitable pseudo-mode amplitudes, based on the poles of the reservoir
structure functions and to show that Markovian equations apply to these
pseudo-mode amplitudes.

The present paper shows (following the approach of Ref.~\cite{Law02}), that
in the case of a three-level atomic system the dynamics is completely
controlled by the reservoir structure functions, and gives several methods
for determining the atomic and field behaviour. These methods could be
applied both to photonic band-gap and high Q cavity cases, since the general
equations (\ref{c32.2}-\ref{c38.3}) defining the solution only depend on the
reservoir structure functions and not on the specific type of structured
reservoir involved. However, as a test, in this paper we only apply the
methods to a situation involving a single Lorentzian reservoir structure
function. This situation could apply when both cascade transitions have the
same frequency and are equally coupled to a single high Q cavity mode. We
also are able to interpret the results via an equivalent pseudo-mode model.
Situations involving photonic band gaps could be modelled by appropriate
reservoir structure functions (see, for example, \cite{Garraway97,Bay98b}).

Section \ref{sec:model} of this paper sets out the theory of non-Markovian
dynamical behaviour for the three level cascade system where both
transitions are coupled to a single structured reservoir. The state
amplitudes are determined from solutions to certain integral equations.
Approaches to solving the dynamical equations, including a numerical
determination of the excited state probability for a simple case (and its
pseudo-mode theory interpretation) is presented in section \ref{sec:solution}%
. Section \ref{sec: separate res} deals with the simpler case of
non-Markovian dynamics for the cascade system with the two transitions
coupled to two separate reservoirs, again with numerical results presented
for comparison to the single reservoir case. An alternative approach to
solving the dynamical equations based on non-orthogonal eigenfunction
methods is set out in Appendices \ref{sec:appA}\textendash\ref{sec:appD}.
The paper is summarised in section \ref{sec:conc}.

\section{Dynamical theory for a single reservoir}

\label{sec:model}


\subsection{The Hamiltonian}

\label{sec:hamil}

The model system we will consider has a three level atom, with states
denoted $|0\rangle $, $|1\rangle $ and $|2\rangle $, coupled to a reservoir
of electromagnetic radiation modes (or heatbath) which is to be at
effectively zero temperature. The bath modes will be described by a density $%
\rho _{\lambda }$, frequency $\omega _{\lambda }$, and raising and lowering
operators $\hat{a}_{\lambda }^{\dagger }$ and $\hat{a}_{\lambda }$.

\begin{figure}[h]
\includegraphics[width=6cm]{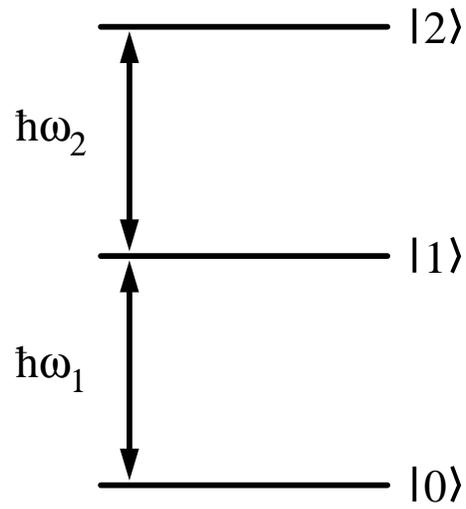}
\caption[setup]{The three level cascade (or ladder) atomic system. The
atomic states 0, 1 and 2 have transition frequencies $\protect\omega_1$ and $%
\protect\omega_2$. }
\label{setup}
\end{figure}

The Hamiltonian for the system is given (in the rotating wave approximation)
by 
\begin{eqnarray}
\hat{H} &=&\hbar \Biggl[\omega _{1}|1\rangle \langle 1|+(\omega _{1}+\omega
_{2})|2\rangle \langle 2|+\sum_{\lambda }\omega _{\lambda }\hat{a}_{\lambda
}^{\dagger }\hat{a}_{\lambda }  \nonumber \\
&&+\sum_{\lambda }\left[ g_{\lambda 1}\left( \hat{a}_{\lambda }^{\dagger
}|0\rangle \langle 1|+\hat{a}_{\lambda }|1\rangle \langle 0|\right)
+g_{\lambda 2}\left( \hat{a}_{\lambda }^{\dagger }|1\rangle \langle 2|+\hat{a%
}_{\lambda }|2\rangle \langle 1|\right) \right] \Biggr] \;,  \label{eq:hamil}
\end{eqnarray}%
where the atomic transition frequencies are $\omega _{1}$ for $%
0\longleftrightarrow 1$ (i.e.\ between atomic states $|0\rangle $ and $%
|1\rangle $) and $\omega _{2}$ for $1\longleftrightarrow 2$ (i.e.\ between
atomic states $|1\rangle $ and $|2\rangle $); see Fig.\ \ref{setup}. The
coupling of electromagnetic radiation with frequency $\omega _{\lambda }$ to
the transition $0\longleftrightarrow 1$ involves the frequency dependent
coupling constant $g_{\lambda 1}$. Likewise, $g_{\lambda 2}$ represents the
coupling of the electromagnetic radiation field to the $1\longleftrightarrow
2$ transition. Real coupling constants will be chosen. It is these frequency
dependent coupling constants combined with the mode density which define the
reservoir structure.


\subsection{Coupled amplitude equations}

\label{sec:analysis}

The Schr\"{o}dinger picture state vector for the three level cascade atom
coupled to the quantum electromagnetic field may be written as: 
\begin{eqnarray}
\left\vert \Psi (t)\right\rangle &=&c_{2}e^{-i(\omega _{1}+\omega
_{2})t}|2\rangle |0_{\lambda }\rangle +\sum_{\lambda }c_{1\lambda
}e^{-i(\omega _{1}+\omega _{\lambda })t}|1\rangle |1_{\lambda }\rangle 
\nonumber \\
&&+\sum_{\lambda }c_{0\lambda \lambda }e^{-2i\omega _{\lambda }t}|0\rangle
|2_{\lambda }\rangle +\sum_{\lambda ,\mu \,,(\lambda <\mu )}c_{0\lambda \mu
}e^{-i(\omega _{\lambda }+\omega _{\mu })t}|0\rangle |1_{\lambda }1_{\mu
}\rangle \;,  \label{c25.1}
\end{eqnarray}%
where $c_{2},c_{1\lambda },c_{0\lambda \lambda }$ and $c_{0\lambda \mu }$
are the amplitudes of the various states in the interaction picture. The
radiation states included are: $|0_{\lambda }\rangle $ in which all the bath
modes are in the vacuum state; $|1_{\lambda }\rangle $ in which the mode
with frequency $\omega _{\lambda }$ has a single excitation, with other
modes being in the vacuum state; $|2_{\lambda }\rangle $ in which the mode
with frequency $\omega _{\lambda }$ has been raised to the second
excitation, with other modes being in the vacuum state; and, $|1_{\lambda
}1_{\mu }\rangle $ in which the modes with frequencies $\omega _{\lambda }$
and $\omega _{\mu }$ have a single excitation, with other modes being in the
vacuum state. A convention for an ordered listing of the modes $\lambda $
for the quantum electromagnetic field is assumed, so that double sum over $%
\lambda ,\mu $ does not lead to a specific state $|1_{\lambda }1_{\mu
}\rangle $ being included twice.

The initial state vector is assumed to be of the form 
\begin{equation}
\left\vert \Psi (0)\right\rangle =|2\rangle |0_{\lambda }\rangle ,
\label{eq:initial_state}
\end{equation}%
which allows us to explore the non-trivial case of two photons appearing in
the reservoir as a result of the interaction with the excited atom.

Substitution of Eq. (\ref{c25.1}) into the time dependent Schr\"{o}dinger
equation gives a closed set of coupled equations for the amplitudes in the
situation where the initial state is given by Eq. (\ref{eq:initial_state}).
If we then take Laplace transforms of the the coupled amplitude equations we
obtain the algebraic equations 
\begin{eqnarray}
s\bar{c}_{2}(s)-1 &=&-i\sum_{\lambda }g_{\lambda 2}\bar{c}_{1\lambda
}(s+i(\omega _{\lambda }-\omega _{2}))  \nonumber \\
s\bar{c}_{1\lambda }(s) &=&-i\sum_{\mu ,(\mu >\lambda )}g_{\mu 1}\bar{c}%
_{0\lambda \mu }(s+i(\omega _{\mu }-\omega _{1}))-i\sum_{\mu ,(\mu <\lambda
)}g_{\mu 1}\bar{c}_{0\mu \lambda }(s+i(\omega _{\mu }-\omega _{1})) 
\nonumber \\
&&-ig_{\lambda 1}\sqrt{2}\bar{c}_{0\lambda \lambda }(s+i(\omega _{\lambda
}-\omega _{1}))-ig_{\lambda 2}\bar{c}_{2}(s+i(\omega _{2}-\omega _{\lambda
}))  \nonumber \\
s\bar{c}_{0\lambda \lambda }(s) &=&-ig_{\lambda 1}\sqrt{2}\bar{c}_{1\lambda
}(s+i(\omega _{1}-\omega _{\lambda }))  \nonumber \\
s\bar{c}_{0\lambda \mu }(s) &=&-ig_{\mu 1}\bar{c}_{1\lambda }(s+i(\omega
_{1}-\omega _{\mu }))-ig_{\lambda 1}\bar{c}_{1\mu }(s+i(\omega _{1}-\omega
_{\lambda }))\quad \quad \quad \quad (\lambda <\mu ),  \label{c25.3}
\end{eqnarray}%
where the Laplace transforms of the amplitudes are $\bar{c}_{2},\bar{c}%
_{1\lambda },\bar{c}_{0\lambda \lambda }$ and $\bar{c}_{0\lambda \mu }$ and
the transform variable is $s$. These equations \ for a cascade system
coupled to a single structured reservoir, are equivalent to those in Ref.\ 
\cite{Nikolopoulos00}. It should be noted that similar equations are given
in Ref.~\cite{Bay98c}, for the case where the two transitions are coupled to
two separate reservoirs. This case is treated below in section \ref{sec:
separate res}. The two separate reservoirs case leads to simpler
equations---firstly, because it is assumed that the $\lambda ,\mu $ photons
are produced in just one sequence (for example $|2\rangle |0_{\lambda
}\rangle \longrightarrow |1\rangle |1_{\lambda }\rangle \longrightarrow
|0\rangle |1_{\lambda }1_{\mu }\rangle $), and secondly, because states of
the form $|0\rangle |2_{\lambda }\rangle $ are not present. Whilst these may
be a good approximation for the single reservoir case when the transition
frequencies are very different, the other process ($|2\rangle |0_{\lambda
}\rangle \longrightarrow |1\rangle |1_{\mu }\rangle \longrightarrow
|0\rangle |1_{\lambda }1_{\mu }\rangle $) would need to also be included
when the transition frequencies are similar, such as in a quantum harmonic
oscillator or a Rydberg atom.

Following the approach of Ref.\ \cite{Law02} we change variables to the
reduced amplitudes $\bar{b}_{2},\bar{b}_{1\lambda },\bar{b}_{0\lambda \mu },%
\bar{b}_{0\lambda \lambda }$ such that 
\begin{eqnarray}
\bar{c}_{2} &=&\bar{b}_{2}  \nonumber \\
\bar{c}_{1\lambda } &=&g_{\lambda 2}\bar{b}_{1\lambda }  \nonumber \\
\bar{c}_{0\lambda \mu } &=&g_{\lambda 2}g_{\mu 1}\bar{b}_{0\lambda \mu
}\quad (\lambda <\mu )  \nonumber \\
\bar{c}_{0\lambda \lambda } &=&g_{\lambda 2}g_{\lambda 1}\bar{b}_{0\lambda
\lambda }\;.  \label{c26.1}
\end{eqnarray}%
Thus we have 
\begin{eqnarray}
s\bar{b}_{2}(s)-1 &=&-i\sum_{\lambda }g_{\lambda 2}^{2}\bar{b}_{1\lambda
}(s+i(\omega _{\lambda }-\omega _{2}))  \label{c27.2a} \\
s\bar{b}_{1\lambda }(s) &=&-i\sum_{\mu ,(\mu >\lambda )}g_{\mu 1}^{2}\bar{b}%
_{0\lambda \mu }(s+i(\omega _{\mu }-\omega _{1}))-i\sum_{\mu ,(\mu <\lambda
)}g_{\mu 1}^{2}\alpha _{\lambda \mu }\bar{b}_{0\mu \lambda }(s+i(\omega
_{\mu }-\omega _{1}))  \label{c27.2b} \\
&&-ig_{\lambda 1}^{2}\sqrt{2}\bar{b}_{0\lambda \lambda }(s+i(\omega
_{\lambda }-\omega _{1}))-i\bar{b}_{2}(s+i(\omega _{2}-\omega _{\lambda })) 
\nonumber \\
s\bar{b}_{0\lambda \lambda }(s) &=&-i\sqrt{2}\bar{b}_{1\lambda }(s+i(\omega
_{1}-\omega _{\lambda }))  \label{c27.2c} \\
s\bar{b}_{0\lambda \mu }(s) &=&-i\bar{b}_{1\lambda }(s+i(\omega _{1}-\omega
_{\mu }))-i\alpha _{\lambda \mu }\bar{b}_{1\mu }(s+i(\omega _{1}-\omega
_{\lambda }))\quad (\lambda <\mu ),  \label{c27.2d}
\end{eqnarray}%
where 
\begin{equation}
\alpha _{\lambda \mu }=\frac{g_{\lambda 1}g_{\mu 2}}{g_{\lambda 2}g_{\mu 1}}%
\;.  \label{c27.2}
\end{equation}%
Analogous equations to (\ref{c27.2a}-\ref{c27.2d}) are given below in Eqs.~(%
\ref{c27.2a:2}-\ref{c27.2d:2}) for the case of a cascade system coupled to
two separate reservoirs.

\subsection{Reservoir structure functions and integral equation to determine
amplitudes}

\label{sec: reservoir structure fns}

Next we eliminate $\bar{b}_{0\lambda \mu },\bar{b}_{0\lambda \lambda }$ by
substitution of (\ref{c27.2c}) and (\ref{c27.2d}) in (\ref{c27.2b}). This
gives 
\begin{eqnarray}
i\bar{b}_{2}(s)+\sum_{\mu }\left[ \left( s+i(\omega _{\lambda }-\omega
_{2})+\sum_{\eta }\frac{g_{\eta 1}^{2}}{s+i(\omega _{\lambda }+\omega _{\eta
}-\omega _{1}-\omega _{2})}\right) \delta _{\lambda \mu }\right. && 
\nonumber \\
\left. +\frac{g_{\mu 1}^{2}\alpha _{\lambda \mu }}{s+i(\omega _{\lambda
}+\omega _{\mu }-\omega _{1}-\omega _{2})}\right] \bar{b}_{1\mu }(s+i(\omega
_{\mu }-\omega _{2})) &=&0\;.  \label{c30.2}
\end{eqnarray}%
Together with Eq. (\ref{c27.2a}) we now have a set of coupled equations for
the $\bar{b}_{2}(s)$ and $\bar{b}_{1\lambda }(s)$. If we eliminate $\bar{b}%
_{2}(s)$ we obtain an equation for the $\bar{b}_{1\lambda }$ alone: 
\begin{eqnarray}
\sum_{\mu }\left[ s\left( s+i(\omega _{\lambda }-\omega _{2})+\sum_{\eta }%
\frac{g_{\eta 1}^{2}}{s+i(\omega _{\lambda }+\omega _{\eta }-\omega
_{1}-\omega _{2})}\right) \delta _{\lambda \mu }\right. &&  \nonumber \\
\left. +s\frac{g_{\mu 1}^{2}\alpha _{\lambda \mu }}{s+i(\omega _{\lambda
}+\omega _{\mu }-\omega _{1}-\omega _{2})}+g_{\mu 2}^{2}\right] \bar{b}%
_{1\mu }(s+i(\omega _{\mu }-\omega _{2})) &=&-i\;.  \label{c30.3}
\end{eqnarray}%
It is useful to rewrite this by by dividing by $s$, using the properties of
the Kronecker delta function, and substituting from Eqs. (\ref{c27.2}) and (%
\ref{c30.3}) to obtain 
\begin{eqnarray}
\left( s+i(\omega _{\lambda }-\omega _{2})+\sum_{\eta }\frac{g_{\eta 1}^{2}}{%
s+i(\omega _{\lambda }+\omega _{\eta }-\omega _{1}-\omega _{2})}\right) \bar{%
b}_{1\lambda }(s+i(\omega _{\lambda }-\omega _{2})) &&  \nonumber \\
+\sum_{\mu }\left( g_{\mu 1}^{2}\alpha _{\lambda \mu }\frac{1}{s+i(\omega
_{\lambda }+\omega _{\mu }-\omega _{1}-\omega _{2})}+\frac{g_{\mu 2}^{2}}{s}%
\right) \bar{b}_{1\mu }(s+i(\omega _{\mu }-\omega _{2})) &=&\frac{-i}{s}\;.
\label{c31.2}
\end{eqnarray}%
We note that the terms involving the quantity $\alpha _{\lambda \mu } $ are
absent in similar equations in Ref.~\cite{Bay98c}, resulting in their
equations for $\bar{b}_{1\lambda }$ being easily solvable. As mentioned
earlier, the additional terms we have result from allowing for photons to be
emitted into the single reservoir in two different sequences, an effect not
present in the two separate reservoir case treated in Ref.~\cite{Bay98c}. In
our case, we next convert the sums to integrals, i.e.\ $\sum_{\mu
}\longrightarrow \int d\omega _{\mu }\rho (\omega _{\mu })$, where $\rho
(\omega _{\mu })$ is the mode density, so that Eq. (\ref{c31.2}) can be
written in the form of an integral equation 
\begin{equation}
A(\omega _{\lambda })\overline{f}(\omega _{\lambda })+\int d\omega _{\mu
}B(\omega _{\lambda },\omega _{\mu })\overline{f}(\omega _{\mu })=C,
\label{c32.2}
\end{equation}%
with 
\begin{eqnarray}
\overline{f}(\omega _{\lambda }) &=&\bar{b}_{1\lambda }(s+i(\omega _{\lambda
}-\omega _{2}))  \label{c32.1.1} \\
A(\omega _{\lambda }) &=&s+i(\omega _{\lambda }-\omega _{2})+\int d\omega
_{\eta }\rho (\omega _{\eta })\frac{g_{\eta 1}^{2}}{s+i(\omega _{\lambda
}+\omega _{\eta }-\omega _{1}-\omega _{2})}  \label{c32.1.2} \\
B(\omega _{\lambda },\omega _{\mu }) &=&\rho (\omega _{\mu })\left( g_{\mu
1}^{2}\frac{g_{\lambda 1}g_{\mu 2}}{g_{\lambda 2}g_{\mu 1}}\frac{1}{%
s+i(\omega _{\lambda }+\omega _{\mu }-\omega _{1}-\omega _{2})}+\frac{g_{\mu
2}^{2}}{s}\right)  \label{c32.1.3} \\
C &=&\frac{-i}{s}\;.  \label{c32.1}
\end{eqnarray}%
The quantities $\overline{f}(\omega _{\lambda }),A(\omega _{\lambda
}),B(\omega _{\lambda },\omega _{\mu })$ and $C$ are of course all functions
of the Laplace variable $s$, but for simplicity of notation $s$ is left
implicit. The integral equation (\ref{c32.2}) is a Fredholm integral
equation of the second kind (see e.g., \cite{Kanwal71}). Methods of solution
for such equations include replacing the frequency spaces by grids of
points, thereby converting the integral equation into matrix equations that
could be solved numerically for each value of $s$. We will discuss one such
approach in section \ref{sec:examp}. In Appendix \ref{sec:appD} we also
discuss a more formal method of solving the integral equation, based on the
eigenfunctions of the kernel $B(\omega _{\lambda },\omega _{\mu })/A(\omega
_{\lambda })$ and of its adjoint.

We also find it convenient to write the integral equation in the form 
\begin{equation}
\overline{f}(\omega _{\lambda })+\int d\omega _{\mu }K(\omega _{\lambda
},\omega _{\mu })\overline{f}(\omega _{\mu })=d(\omega _{\lambda }),
\label{c33.2}
\end{equation}%
where 
\begin{eqnarray}
d(\omega _{\lambda }) &=&C/A(\omega _{\lambda })  \label{c33.3.1} \\
K(\omega _{\lambda },\omega _{\mu }) &=&B(\omega _{\lambda },\omega _{\mu
})/A(\omega _{\lambda })\;.  \label{c33.3}
\end{eqnarray}

We note that the coupling constants and mode density appear in the integral
equation only in the form \textquotedblleft $\rho g^{2}$\textquotedblright .
These forms are called reservoir structure functions, as they contain all
the essential features of the reservoir and its coupling to the atomic
system. Specifically, the reservoir structure functions that appear in Eqs. (%
\ref{c32.1.2}) and (\ref{c32.1.3}) are 
\begin{eqnarray}
R_{1}(\omega _{\lambda }) &=&\rho (\omega _{\lambda })g_{\lambda 1}^{2}, 
\nonumber \\
R_{2}(\omega _{\lambda }) &=&\rho (\omega _{\lambda })g_{\lambda 2}^{2}.
\label{eq.res.struc.fns}
\end{eqnarray}%
As the coupling constants are proportional to dipole matrix elements
multiplied by the square root of the angular frequency, it is clear that the
factors $g_{\mu 2}/g_{\mu 1}$ and $g_{\lambda 1}/g_{\lambda 2}$ in Eq. (\ref%
{c32.1.3}) are independent of the frequencies $\omega _{\lambda }$ and $%
\omega _{\mu }$. Hence a third reservoir structure function involving the
factor $\alpha _{\lambda \mu }$ is not needed. As the dipole matrix elements
would essentially cancel out, the factor $\alpha _{\lambda \mu }$ is of
order unity.

In principle, we can solve the integral equation and thus determine the $%
\bar{b}_{1\lambda }(s+i(\omega _{\lambda }-\omega _{2}))$. Furthermore, the
solutions obtain their particular form from just the reservoir structure
functions, rather than the density of states or coupling constants alone.

Next we see that in the new notation Eq.~(\ref{c27.2a}) becomes 
\begin{eqnarray}
\bar{b}_{2}(s) &=&\frac{1}{s}-\frac{i}{s}\sum_{\lambda }g_{\lambda 2}^{2}%
\bar{b}_{1\lambda }(s+i(\omega _{\lambda }-\omega _{2}))  \nonumber \\
&\equiv &\frac{1}{s}-\frac{i}{s}\int d\omega _{\lambda }\rho (\omega
_{\lambda })g_{\lambda 2}^{2}\overline{f}(\omega _{\lambda }) ,
\label{c38.3}
\end{eqnarray}%
and again the step to obtaining $\bar{b}_{2}(s)$ just involves using the
reservoir structure function $R_{2}(\omega _{\lambda })$. Note again that $%
\overline{f}(\omega _{\lambda })$ is a function of the Laplace variable $s$,
so the decay of the initial atomic state $|2\rangle $ described by $\bar{b}%
_{2}(s)$ is non-exponential in general.

Finally, we note Eqs. (\ref{c27.2c}) and (\ref{c27.2d}) imply that $\bar{b}%
_{0\lambda \mu }$ and $\bar{b}_{0\lambda \lambda }$ are fully determined
once $\bar{b}_{2},\bar{b}_{1\lambda }$ are known, and $\alpha _{\lambda \mu
} $ (Eq. (\ref{c27.2})) introduces no new frequency dependence as it is
independent of frequency. Thus all the reduced amplitudes $\bar{b}_{2},\bar{b%
}_{1\lambda },\bar{b}_{0\lambda \mu }$, and $\bar{b}_{0\lambda \lambda }$
can be determined in principle from reservoir structure functions. As we
will see next, this is sufficient to determine the reduced density operator
describing the atomic system.

Note that the non-Markovian methods could be applied both to photonic
band-gap and high Q cavity cases, since the general equations (\ref{c32.2}-%
\ref{c38.3}) defining the solution only depend on the reservoir structure
functions and not on the specific type of structured reservoir involved. 
Markovian results can be obtained under conditions where the reservoir
structure functions $\rho (\omega _{\lambda })g_{\lambda 1,\lambda 2}^{2}$
are slowly varying functions of $\omega _{\lambda }$. Certain
integrals give a constant term whose imaginary parts are the (formally 
divergent) frequency shifts and whose real parts are the decay rates for the
states $|1\rangle$ and $|2\rangle $.

\subsection{Atomic density operator}

\label{sec:atomic density}

The atomic density operator is defined by 
\begin{equation}
\hat{\rho}_{A}=\mbox{Tr}_{F}\left\vert \Psi \right\rangle \left\langle \Psi
\right\vert ,  \label{eq:density_op_atomic}
\end{equation}%
and it is not difficult to show that 
\begin{eqnarray}
\hat{\rho}_{A} &=&\left\vert b_{2}(t)\right\vert ^{2}|2\rangle \langle
2|+\left( \int d\omega _{\lambda }\rho (\omega _{\lambda })g_{\lambda
2}^{2}\left\vert b_{1\lambda }(t)\right\vert ^{2}\right) |1\rangle \langle 1|
\nonumber \\
&&+\left( \int \int_{\lambda \leq \mu }d\omega _{\lambda }d\omega _{\mu
}\rho (\omega _{\lambda })\rho (\omega _{\mu })g_{\lambda 2}^{2}g_{\mu
1}^{2}\left\vert b_{0\lambda \mu }(t)\right\vert ^{2}\right) |0\rangle
\langle 0|\;.  \label{eq:density_op_atomic_expression}
\end{eqnarray}%
Thus we see that the atomic operator only depends on the reduced amplitudes $%
b_{2}(t),b_{1\lambda }(t),b_{0\lambda \mu }(t)\,(\lambda \leq \mu )$, and
the reservoir structure functions. As the former can be determined, in
principle, from the reservoir structure functions, we see that the behaviour
of the cascade atom in the structured reservoir is completely determined by
the reservoir structure functions (for the initial state given in Eq. (\ref%
{eq:initial_state})).

On the basis of this key result, it would follow that any existing system
could be replaced by an equivalent system, provided that the reservoir
structure functions were the same in both models. This is the basis of the
treatment of superradiance in a photonic band gap continuum \cite{Bay98b},
where the photonic band gap system is replaced by a pair of degenerate
cavity modes coupled to the multi-atom system and with each other, one of
the modes being also coupled to a Markovian bath. In terms of the treatment
in \cite{Dalton01a}, such a case would produce the required Fano-profile
reservoir structure function, with the Fano window representing the photonic
band gap. The two cavity modes would correspond to two pseudo-modes.

The absence of any coherence terms in the atomic density operator is a
consequence of the choice of initial state, Eq. (\ref{eq:initial_state}).
The choice of a more general initial state (even with no photons present) of
the form%
\begin{equation}
\left\vert \Psi (0)\right\rangle =(c_{2}|2\rangle +c_{1}|1\rangle
+c_{0}|0\rangle )|0_{\lambda }\rangle
\end{equation}%
would require the introduction of a more general time dependent state vector 
$\left\vert \Psi (t)\right\rangle $ than that given in Eq. (\ref{c25.1}), to
include additional states of the form $|0\rangle |0_{\lambda }\rangle
,|1\rangle |0_{\lambda }\rangle $ and $|0\rangle |1_{\lambda }\rangle $. The
amplitudes for these additional states are not coupled to those for the
other states included in Eq. (\ref{c25.1}). Again, the solutions for these
amplitudes just involve reservoir structure functions and are analogous to
those already discussed in Ref.~\cite{Garraway97} for the simpler case of a
two level atom coupled to a structured reservoir. However, as indicated
above, the atomic density operator would then include coherence terms.

\section{Solutions for the state amplitudes}

\label{sec:solution}

The integral equation (\ref{c32.2}) can be solved in different ways. These
include: (a) numerical methods based on converting the integral equation to
a matrix equation, (b) expansions using biorthogonal eigenfunctions and (c)
expansions such as the Fredholm expansion \cite{Kanwal71}. Only the first of
these methods will be used here, but as the second approach using
biorthogonal eigenfunctions may be used in later work and has not been used
previously in Quantum Optics problems, it is included here in Appendices \ref%
{sec:appA} - \ref{sec:appD} for completeness.

\subsection{Numerical solution of the integral equation: case of Lorentzian
reservoir structure function}

\label{sec:examp}

As an illustration we consider a greatly simplified example of a three-level
system coupled to a reservoir with structure. The simplest possible case is
that for the same Lorentzian reservoir structure function associated with
both transitions, with all the couplings and transition frequencies equal to
each other. That is, we choose a single coupling constant $g_{\lambda }$
such that 
\begin{equation}
g_{\lambda 1}=g_{\lambda 2}=g_{\lambda },  \label{example:g}
\end{equation}%
which amounts to both the dipole moment matrix elements for the transitions
being equal. The atom will also have two equally spaced transitions which
are resonant with the reservoir structure, 
\begin{equation}
\omega _{1}=\omega _{2}=\omega _{0}\;.  \label{example:omega}
\end{equation}%
We refer only to $\omega _{0}$ in the following. Thus for the single
reservoir structure function, we have 
\begin{equation}
R_{1}=R_{2}=\rho _{\lambda }g_{\lambda }^{2}=\frac{\Gamma \Omega ^{2}}{2\pi }%
\,\cdot \,\frac{1}{(\omega _{\lambda }-\omega _{0})^{2}+(\Gamma /2)^{2}}
\label{example:rhogsq}
\end{equation}%
as in Ref.\ \cite{Garraway97}. The parameter $\Omega $ represents the
strength of the coupling and $\Gamma $ represents the width of the reservoir
structure function. This situation would apply for identical cascade
transitions coupled to a single high Q cavity mode. Cascade transitions in a
photonic band gap reservoir would be treated via a different choice of the
reservoir structure functions.

Using this expression for the reservoir structure function we can determine
the functions $A(\omega _{\lambda }),B(\omega _{\lambda },\omega _{\mu })$
and $C$ in Eqs. (\ref{c32.1.2}-\ref{c32.1}) and then the kernel, Eq. (\ref%
{c33.3}), becomes 
\begin{equation}
K(\omega _{\lambda },\omega _{\mu })=\frac{\Gamma \Omega ^{2}}{2\pi }\,\cdot
\,\frac{(s+i(\omega _{\lambda }-\omega _{0})+\Gamma /2)(2s+i(\omega
_{\lambda }+\omega _{\mu }-2\omega _{0}))}{s((\omega _{\mu }-\omega
_{0})^{2}+(\Gamma /2)^{2})(s+i(\omega _{\lambda }+\omega _{\mu }-2\omega
_{0}))Q(\omega _{\lambda }-\omega _{0})},  \label{example:kernel}
\end{equation}%
where $Q(\omega )$ is a quadratic polynomial such that 
\begin{equation}
Q(\omega )=(s+i\omega )(s+i\omega +\Gamma /2)+\Omega ^{2}\;.
\end{equation}

For this model we thus have an analytic form for the kernel, but to go
further it appears that we need to use a numerical method. We could utilize
an eigenfunction method, such as that of Appendix \ref{sec:appD}, but choose
a very simple approach to solve Eq. (\ref{c33.2}). The process is simply to
represent Eq. (\ref{c33.2}) as a matrix equation 
\begin{equation}
(\mathbf{K}+\mathbf{I})\overline{\mathbf{f}}=\mathbf{d}\;,
\label{matrix_eqn}
\end{equation}%
where $\mathbf{K}$ and $\mathbf{I}$ are matrices and $\overline{\mathbf{f}}$
and $\mathbf{d}$ are vectors. We then invert $(\mathbf{K}+\mathbf{I})$ to
solve for $\overline{\mathbf{f}}$. Thus $K(\omega _{\lambda },\omega _{\mu
}) $ is represented at discrete frequency points, in effect a discrete basis
of spatial delta-functions, e.g., $\mathbf{K}_{\omega _{\lambda },\omega
_{\mu }}=K(\omega _{\lambda },\omega _{\mu })$. Similarly, $\overline{f}%
(\omega _{\lambda })$ and $d(\omega _{\lambda })$ are represented at
discrete frequency points. From the definition in Eq. (\ref{c32.1.1}), we
see that if we introduce the function $f(\omega _{\lambda },t)$ (which we
denote as $f(t) $ for short) via 
\begin{equation}
f(\omega _{\lambda },t)=\exp [-i(\omega _{\lambda }-\omega
_{2})t]\;b_{1\lambda }(t),
\end{equation}%
then $f(t)$ is the function whose Laplace transform is $\overline{f}%
(s)\equiv \overline{f}(\omega _{\lambda },s)$. However, in order to obtain
the real and imaginary parts of $f(t)$, we will need the separate inverse
Laplace transforms $\overline{f_{r}}(s),\overline{f_{i}}(s)$ which are the
Laplace transforms of the real and imaginary parts of $%
f(t)=f_{r}(t)+if_{i}(t)$. For complex $s$ the latter Laplace transforms 
\textit{cannot} be obtained by just writing $\overline{f}(s)$ as the sum of
its real and imaginary parts. However, the Laplace transform $\overline{f_{r}%
}(s)$ of the real part $f_{r}(t)$ is real (and similarly the Laplace
transform $\overline{f_{i}}(s)$ of the imaginary part $f_{i}(t)$ is real),
if the Laplace transform parameter $s$ is real. Hence, the real and
imaginary parts of $\overline{f}(s)$ are equal to the Laplace transforms of
the real and imaginary parts of $f(t)$ for $s$ on the real axis, so $%
\mbox{Re}\overline{f}(s)=\overline{f_{r}}(s),\mbox{Im}\overline{f}(s)=%
\overline{f_{i}}(s)$ for $s$ real. As $\overline{f}(s)$ is an analytic
function of $s$, the analytic continuation of $\overline{f_{r}}(s)+i%
\overline{f_{i}}(s)$ from the real axis will determine $\overline{f}(s)$
everywhere.

In this example, if we discretise $K(\omega _{\lambda },\omega _{\mu })$ on
an $N\times N$ grid we define the $N\times N$ complex matrices $\mathbf{K}%
_{r}$ and $\mathbf{K}_{i}$ from the real and imaginary parts of Eq.~(\ref%
{example:kernel}) on the real $s$-axis, and then Eq.~(\ref{matrix_eqn})
becomes 
\begin{equation}
\left( 
\begin{array}{c|c}
\mathbf{K}_{r}+\mathbf{I} & -\mathbf{K}_{i} \\ \hline
\mathbf{K}_{i} & \mathbf{K}_{r}+\mathbf{I}%
\end{array}%
\right) \left( 
\begin{array}{c}
\overline{\mathbf{f}_{r}} \\ \hline
\overline{\mathbf{f}_{i}}%
\end{array}%
\right) =\left( 
\begin{array}{c}
\mathbf{d}_{r} \\ \hline
\mathbf{d}_{i}%
\end{array}%
\right) ,  \label{matrix_eqn2}
\end{equation}%
for $s$ on the real axis. The formal solution for $\overline{\mathbf{f}_{r}}=%
\overline{\mathbf{f}_{r}}(s)$ and $\overline{\mathbf{f}_{i}}=\overline{%
\mathbf{f}_{i}}(s)$ for real $s$ then generates the solution for $\overline{%
\mathbf{f}}(s)$ everywhere. Because of this (and having first identified $%
\mathbf{K}_{r},\mathbf{K}_{i},$ $\mathbf{d}_{r}$ and $\mathbf{d}_{i}$ for
real $s$ using Eq. (\ref{c33.3}) and Eq. (\ref{c33.3.1})), we can \textit{%
now }regard Eq. (\ref{matrix_eqn2}) as applying for \textit{all }$s$. This
approach could not be used if the the real and imaginary parts of $K(\omega
_{\lambda },\omega _{\mu })$ and $d(\omega _{\lambda })$ on the real $s$%
-axis are not analytic. The matrix inversion step thus involves a matrix
with $4N^{2}$ elements compared to, say, $\mathcal{O}(N^{4})$ elements
represented by Eqs. (\ref{c27.2a} - \ref{c27.2d}) in an equivalent
discretised form.

Thus we solve for $\overline{\mathbf{f}_{r}}$ and $\overline{\mathbf{f}_{i}}$
in Eq. (\ref{matrix_eqn2}), and hence determine the $\bar{b}_{1\lambda
}(s+i(\omega _{\lambda }-\omega _{2}))$ of Eq. (\ref{c32.1.1}). We then find
the $\bar{b}_{2}(s)$ from the scalar product form Eq.~(\ref{c38.3}) obtained
from Eq. (\ref{c27.2a}) so that 
\begin{eqnarray}
\overline{b}_{2r}(s) &=&(1+\mathbf{r\cdot }\overline{\mathbf{f}_{i}})/s 
\nonumber \\
\overline{b}_{2r}(s) &=&-\mathbf{r\cdot }\overline{\mathbf{f}_{r}}/s,
\label{eq.scalprod}
\end{eqnarray}%
where $\mathbf{r}\equiv \{\rho _{\lambda }g_{\lambda 2}^{2}\}$. Finally, $%
b_{2}(t)$ is determined by a numerical inverse Laplace transform.

\begin{figure}[tbp]
\includegraphics{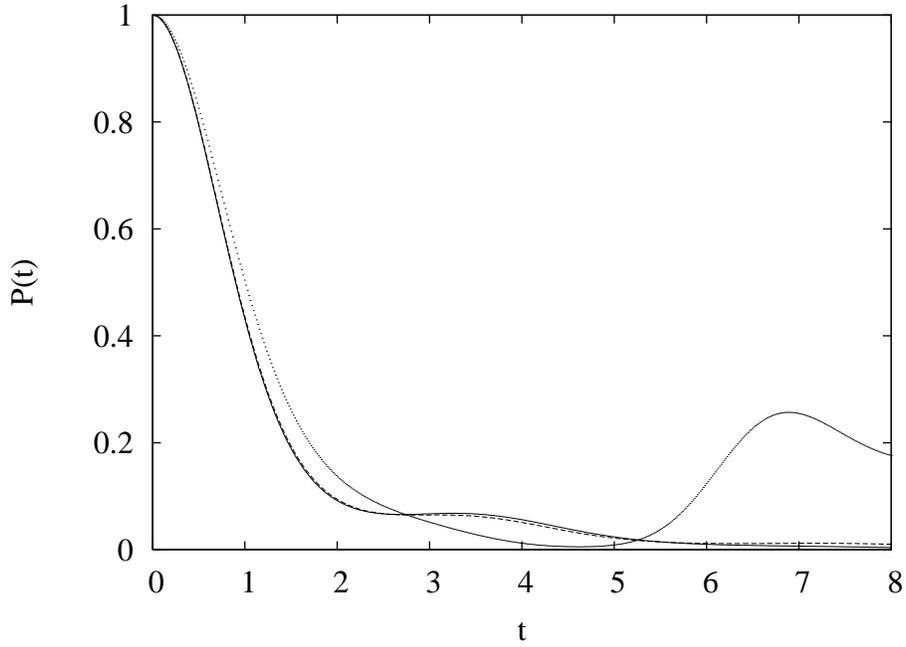}
\caption[fig2]{Time evolution of the probability of finding the system in
state 2; $P(t)=|b_{2}(t)|^{2}$. The reservoir structure function is given by
Eq.~(\protect\ref{example:rhogsq}) with $\Gamma $ =1 and $\Omega $ = 1 in
scaled units. The grid size for the discretised kernel was: 150$\times 150$
(solid), 100$\times 100$ (dashed), and 50$\times 50$ (dotted). In each case
a range of $\pm $30 for $\protect\omega_{\protect\lambda} -\protect\omega %
_{0}$ and $\protect\omega_{\protect\mu} -\protect\omega _{0}$ was chosen.
The result for a grid size of 150$\times 150$ (solid curve) gives a
reasonably accurate result. }
\label{fig2}
\end{figure}

Figure~\ref{fig2} shows some results for this numerical matrix approach with
the kernel given in~Eq. (\ref{example:kernel}), which was derived from the
reservoir structure function in Eq. (\ref{example:rhogsq}). The three curves
show the upper state population for three different sizes of matrix which
were used to discretize the integral equation. Each case used the same
parameters $\Omega $=1 and $\Gamma $=1, where there is a distinct
non-Markovian evolution that could not be treated perturbatively because of
the strong coupling to the reservoir structure. The solid curve in Fig.~\ref%
{fig2} shows a good result that was obtained with a matrix of size 150$%
\times 150$ for this problem. Reducing the matrix size to 100$\times 100$
(dashed) results in only a slight degradation of the result. However,
further reduction of the matrix size affects the numerical result quite
badly.

\begin{figure}[tbp]
\includegraphics{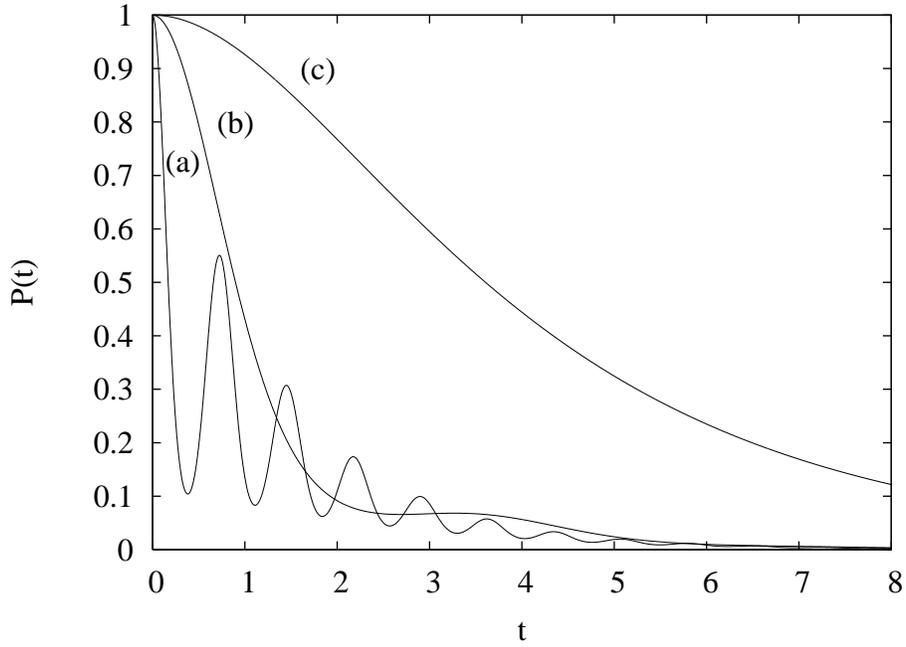}
\caption[fig3]{Time evolution of the probability of finding the system in
state 2; $P(t)=|b_{2}(t)|^{2}$. The reservoir structure function is Eq.~(%
\protect\ref{example:rhogsq}) with $\Gamma $ =1 and: (a) $\Omega $ = 5.0;
(b) $\Omega $ = 1.0; and (c) $\Omega $ = 0.3, in scaled units. The grid size
for the discretised kernel was 150$\times 150$ chosen with a range of $\pm $%
30 for $\protect\omega_{\protect\lambda} -\protect\omega _{0}$ and $\protect%
\omega _{\protect\mu} -\protect\omega_{0}$ in scaled units (as in Fig.~%
\protect\ref{fig2}).}
\label{fig3}
\end{figure}

The effect of changing the coupling strength $\Omega $ is shown in Fig.~\ref%
{fig3}. The probability of finding the atomic system in the highest atomic
state is shown. For strong coupling (Fig.~\ref{fig3}(a)) we see damped
oscillations that are a typical manifestation of non-Markovian processes. As
the coupling is reduced, (Fig.~\ref{fig3}(b)), the oscillations weaken and
then further reductions in the coupling strength $\Omega $, (Fig.~\ref{fig3}%
(c)) result in no oscillations and decay that is closer to exponential and
on a longer time-scale than the strong coupling cases. Fig.~\ref{fig3}(c)
still shows some visible initial quadratic behaviour because of the
relatively high value of $\Omega /\Gamma $.

\subsection{Equivalent pseudo-mode model}

The reservoir structure function given in Eq.~(\ref{example:rhogsq}) is
extremely simple and as result we can reproduce the results of Fig.~\ref%
{fig3}, i.e.\ the population $|b_{2}(t)|^{2}$, from the Markovian master
equation 
\begin{equation}
\frac{\partial \hat{\rho}}{\partial t}=-i[\hat{V},\hat{\rho}]-\frac{\Gamma }{%
2}\left( \hat{a}^{\dagger }\hat{a}\hat{\rho}+\hat{\rho}\hat{a}^{\dagger }%
\hat{a}\hat{\rho}-2\hat{a}\hat{\rho}\hat{a}^{\dagger }\right) ,
\label{eq:me}
\end{equation}%
which is given in the interaction picture with the atom-`field' coupling
term 
\begin{equation}
\hat{V}=\Omega \left( \hat{a}^{\dagger }|0\rangle \langle 1|+\hat{a}%
|1\rangle \langle 0|+\hat{a}^{\dagger }|1\rangle \langle 2|+\hat{a}|2\rangle
\langle 1|\right) \;.  \label{eq:meH}
\end{equation}%
In this master equation we have introduced a single oscillator, or
pseudo-mode \cite{Garraway97}, which is represented by the harmonic
oscillator operators $\hat{a}$ and $\hat{a}^{\dagger }$. In this approach
(see \cite{Garraway97}) pseudo-modes are introduced as assumed bosonic
entities, rather than via constructing pseudo-mode amplitudes. A cascade
atom resonantly coupled to a damped high-Q cavity mode, which is also
coupled to a Markovian bath of vacuum modes, is an example of a physical
system which has the same master equation as (\ref{eq:me}). Such a model was
considered in our earlier work \cite{Dalton01a}, where we showed that
multiple excitations of a structured reservoir could be treated for
reservoir structure functions such as Eq. (\ref{example:rhogsq}). To utilize
the present pseudo-mode model we solve the master equation (\ref{eq:me})
with the initial condition of an empty pseudo-mode and the atom in the state 
$|2\rangle $. On tracing out the pseudo-mode, to obtain atomic properties
alone, we can reproduce the results of the matrix method used with the
kernel of Eq. (\ref{example:kernel}). It should be emphasised that it does
not appear to be easy to find such a simple master equation for more complex
reservoir structures such as photonic band gap models with branch cuts in
the reservoir structure function. In such a case the approach outlined in
this paper (which only depends on the reservoir structure functions) may be
useful instead. For the present Lorentzian model, the agreement between the
matrix method given earlier in this section and the master equation (\ref%
{eq:me}) is excellent.


\section{Dynamical theory for two separate reservoirs}

\label{sec: separate res}

In this paper we have commented in several places that there are differences
in our single reservoir treatment from the simpler case of separate
reservoirs coupled to the two transitions in our model system. In our model,
the two photons may be emitted in either order, whereas with the
distinguishable photons in the two reservoir model, only one order of
emission is involved. So, with the formalism now complete, it is instructive
to look at the explicit differences between our model and the simpler two
separate reservoirs model of the kind considered in Ref. \cite{Bay98c}. In
this case the Hamiltonian in Eq. (\ref{eq:hamil}) is replaced by 
\begin{eqnarray}
\hat{H} &=&\hbar \Biggl[\omega _{1}|1\rangle \langle 1|+(\omega _{1}+\omega
_{2})|2\rangle \langle 2|+\sum_{\lambda }\omega _{\lambda }\hat{a}_{\lambda
}^{\dagger }\hat{a}_{\lambda }+\sum_{\mu }\omega _{\mu }\hat{b}_{\mu
}^{\dagger }\hat{b}_{\mu }  \nonumber \\
&&+\sum_{\mu }g_{\mu 1}\left( \hat{b}_{\mu }^{\dagger }|0\rangle \langle 1|+%
\hat{b}_{\mu }|1\rangle \langle 0|\right) +\sum_{\lambda }g_{\lambda
2}\left( \hat{a}_{\lambda }^{\dagger }|1\rangle \langle 2|+\hat{a}_{\lambda
}|2\rangle \langle 1|\right) \Biggr]\;,  \label{eq:hamil2}
\end{eqnarray}%
where the bath operators $\hat{a}_{\lambda }^{\dagger }$ and $\hat{a}%
_{\lambda }$ for the first bath now couple only to the $1\longleftrightarrow
2$ transition, and the new bath operators $\hat{b}_{\mu }^{\dagger }$ and $%
\hat{b}_{\mu }$ \ for the second bath couple only to the $%
0\longleftrightarrow 1$ transition. For the initial state vector, Eq. (\ref%
{eq:initial_state}), the state vector analogous to Eq. (\ref{c25.1}) no
longer contains a term involving $c_{0\lambda \lambda }$, and there is no
restriction over the double sum $\lambda ,\mu $, since the two types of bath
modes are now distinct. We can write 
\begin{eqnarray}
\left\vert \Psi (t)\right\rangle &=&c_{2}e^{-i(\omega _{1}+\omega
_{2})t}|2\rangle |0_{\lambda }\rangle |0_{\mu }\rangle +\sum_{\lambda
}c_{1\lambda }e^{-i(\omega _{1}+\omega _{\lambda })t}|1\rangle |1_{\lambda
}\rangle |0_{\mu }\rangle  \nonumber \\
&&+\sum_{\lambda ,\mu }c_{0\lambda \mu }e^{-i(\omega _{\lambda }+\omega
_{\mu })t}|0\rangle |1_{\lambda }\rangle |1_{\mu }\rangle \;,
\label{c25.1:2}
\end{eqnarray}%
involving product states of: the atom, and one or zero excitation states of
the two baths. The equations for the Laplace transforms of the reduced
amplitudes, Eqs. (\ref{c27.2a}-\ref{c27.2d}) are then replaced by 
\begin{eqnarray}
s\bar{b}_{2}(s)-1 &=&-i\sum_{\lambda }g_{\lambda 2}^{2}\bar{b}_{1\lambda
}(s+i(\omega _{\lambda }-\omega _{2}))  \label{c27.2a:2} \\
s\bar{b}_{1\lambda }(s) &=&-i\sum_{\mu }g_{\mu 1}^{2}\bar{b}_{0\lambda \mu
}(s+i(\omega _{\mu }-\omega _{1}))-i\bar{b}_{2}(s+i(\omega _{2}-\omega
_{\lambda }))  \label{c27.2b:2} \\
s\bar{b}_{0\lambda \mu }(s) &=&-i\bar{b}_{1\lambda }(s+i(\omega _{1}-\omega
_{\mu })),  \label{c27.2d:2}
\end{eqnarray}%
We note that at this point the differences are that, as well as the absence
of the $\bar{b}_{0\lambda \lambda }$ terms, there are no terms involving $%
\alpha _{\lambda \mu }$ (as in Eqs. (\ref{c27.2b}) and (\ref{c27.2d})) and
there are no restrictions over the sum over $\mu $ (as in Eq. (\ref{c27.2b}%
)) These equations are equivalent to those in Ref. \cite{Bay98c}.

As in the case of both transitions coupled to one single reservoir, the
dynamical behaviour only depends on reservoir structure functions, and
following the same approach as in section \ref{sec:atomic density} it is
easy to see that the atomic density operator is also determined from these
functions.

If we now follow the elimination procedure of section \ref{sec: reservoir
structure fns} we find the same equations (\ref{c32.2}-\ref{c32.1}) for $%
\overline{f}$, $A$, $B$ and $C$ except that the consequence of no $\alpha
_{\lambda \mu }$ term being present in Eq. (\ref{c27.2d:2}) is that the
quantity $B$ becomes 
\begin{equation}
B(\omega _{\lambda },\omega _{\mu })\longrightarrow B(\omega _{\mu })=\rho
(\omega _{\mu })\frac{g_{\mu 2}^{2}}{s}
\end{equation}%
Crucially $B$ no longer depends on $\omega _{\lambda }$ as previously.
Expressions for $\overline{f}$, $A$ and $C$ are otherwise unchanged.

The integral equation then simplifies to the easily solvable form 
\begin{equation}
A(\omega _{\lambda })\overline{f}(\omega _{\lambda })+\int d\omega _{\mu
}B(\omega _{\mu })\overline{f}(\omega _{\mu })=C,  \label{c32.2:2}
\end{equation}%
for which the solution is 
\begin{equation}
\overline{f}(\omega _{\lambda })=\frac{C}{1+\int d\omega _{\mu }K(\omega
_{\mu },\omega _{\mu })}\cdot \frac{1}{A(\omega _{\lambda })}.
\label{eq: f solution}
\end{equation}%
In this case the equivalent kernel is separable: $K(\omega _{\lambda
},\omega _{\mu })=B(\omega _{\mu })/A(\omega _{\lambda })$.

We can apply our results to the situation analogous to that treated in
section \ref{sec:examp}, where both reservoirs, though now separate, have
identical coupling constants and reservoir structure functions, and the two
atomic transitions are equally spaced and resonant with the reservoir
structures. We utilize Eqs.\ (\ref{example:g}-\ref{example:rhogsq}) and, for
this simple model, the kernel can be easily obtained as: 
\begin{equation}
K(\omega _{\lambda },\omega _{\mu })=\frac{\Gamma \Omega ^{2}}{2\pi }\,\cdot
\,\frac{(s+i(\omega _{\lambda }-\omega _{0})+\Gamma /2)}{s((\omega _{\mu
}-\omega _{0})^{2}+(\Gamma /2)^{2})Q(\omega _{\lambda }-\omega _{0})}.
\label{example:kernel2}
\end{equation}%
This result may be compared to the previous expression in Eq. (\ref%
{example:kernel}) for the case of a single reservoir.

The integral $\int d\omega _{\mu }K(\omega _{\mu },\omega _{\mu })$ can be
performed by using a contour in the lower-half plane, and we obtain 
\begin{equation}
\int d\omega _{\mu }K(\omega _{\mu },\omega _{\mu })=\frac{\Omega ^{2}}{s}%
\frac{s+\Gamma }{(s+\Gamma /2)(s+\Gamma )+\Omega ^{2}} .
\end{equation}%
We may now find from Eq. (\ref{c32.1.2}) that 
\begin{equation}
A(\omega _{\lambda })=s+i(\omega _{\lambda }-\omega _{0})+\frac{\Omega ^{2}}{%
s+\Gamma /2+i(\omega _{\lambda }-\omega _{0})}  \label{eq:2resA}
\end{equation}%
so the solution for $\overline{f}(\omega _{\lambda })$ can be obtained from
Eq. (\ref{eq: f solution}). We find that: 
\begin{equation}
\overline{f}(\omega _{\lambda })= -i \frac{\left[(s+\Gamma /2)(s+\Gamma
)+\Omega ^{2}\right] \left[ s+ i(\omega _{\lambda }-\omega _{0})+\Gamma /2 %
\right]}{ (s+\Gamma /2) \left[ s (s+\Gamma) +2\Omega^2 \right] Q(\omega
_{\lambda }-\omega_{0})}.  \label{eq: f result}
\end{equation}

A numerical inversion of $\overline{f}(\omega _{\lambda })$ can be performed
to obtain $b_{2}(t)$ using the same approach as in section \ref{sec:examp}.
However, The reservoir structure, Eq.~(\ref{example:rhogsq}), is
sufficiently simple that a solution for $b_{2}(t)$ can be found from Eq.~(%
\ref{eq: f result}). We first need to perform the integral in Eq.~(\ref%
{c38.3}) which is facilitated by the fact that Eq.~(\ref{eq: f result}) has
no poles in the lower-half complex plane (for Re$(s) > 0$), while the factor 
$\rho (\omega_{\lambda })g_{\lambda 2}^{2}$ in Eq.~(\ref{c38.3}) has only a
single pole in the lower-half complex plane if we use the example given in
Eq.~(\ref{example:rhogsq}). Then if we perform the integral of  Eq.~(\ref%
{c38.3}) around the single lower-half plane pole we find that  
\begin{equation}
b_{2}(s) = \frac{1}{s} -\Omega^2 \frac{ s + \Gamma}{ s (s+\Gamma /2) \left[
s (s+\Gamma) +2\Omega^2 \right] }  
\, .
\label{eq:2res_b2s}
\end{equation}
If we now perform the inverse Laplace transform, we find 
\begin{equation}
b_{2}(t) = \frac{ \Omega^2 }{\beta^2} e^{-\Gamma t / 2} + \left( 1 - \frac{
\Omega^2 }{\beta^2} \right) e^{-\Gamma t / 2} \cos(\beta t ) + \frac{ \Gamma 
}{2 \beta} e^{-\Gamma t / 2} \sin(\beta t )  
\, ,
\label{eq:2res_b2t}
\end{equation}
where $\beta^2 = 2\Omega^2 - (\Gamma/2)^2$.

\begin{figure}[tbp]
\includegraphics{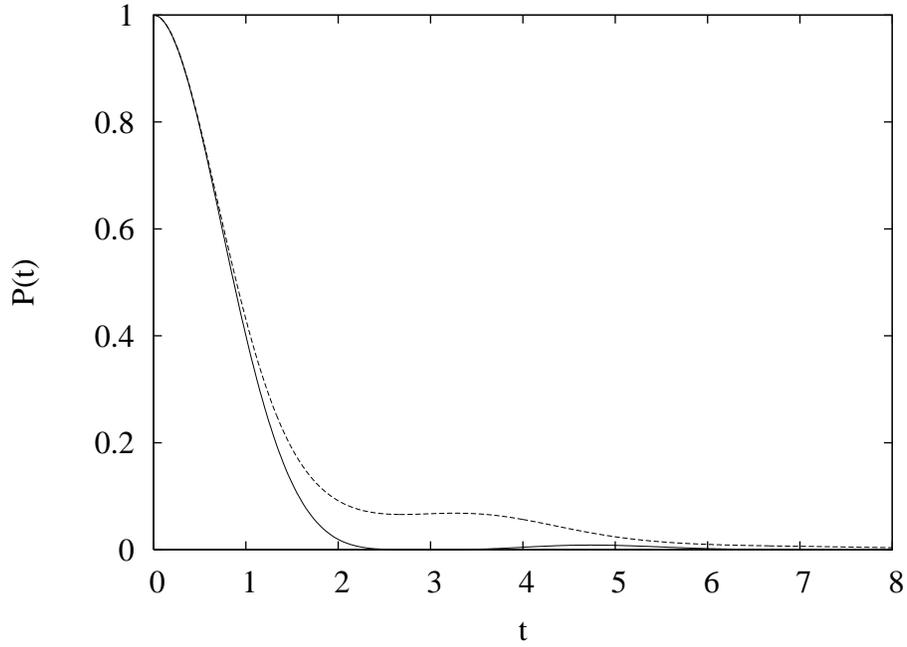}
\caption[fig4]{Time evolution of the probability of finding the system in
state 2; $P(t)=|b_{2}(t)|^{2}$. The reservoir structure function is Eq.~(%
\protect\ref{example:rhogsq}) with $\Gamma $ =1 and $\Omega $ = 1.0. The two
curves show the effect of changing from two separate reservoirs (solid line)
to a single reservoir (dashed line). (Other parameters are as given in Fig.\ 
\protect\ref{fig2}. The dashed line in this figure is identical to the solid
line in Fig.\ \protect\ref{fig2}.) }
\label{fig4}
\end{figure}

The result for the time evolution of the probability for finding the atom in
the highest atomic state is seen in Fig.\ \ref{fig4}. There is clearly a
difference from the single reservoir result shown in Fig.\ \ref{fig3} (the
dashed line in Fig.\ \ref{fig4}). The present situation, where both atomic
transition frequencies are equal and resonant with the structured reservoir,
should highlight the difference between the cases of two separate or one
single reservoir. In this situation both photons emitted should have similar
frequencies, and the single reservoir case where the first emitted photon
cannot be distinguished from the other should give different results to the
two distinct reservoir case where they can.

We note that for strongly coupled systems, $2\Omega^2 > (\Gamma/2)^2$, the
time evolution in Eq.~(\ref{eq:2res_b2t}) can be re-expressed in the form 
\begin{equation}
b_{2}(t) = \frac{ 2\Omega^2 }{ 2\Omega^2 - (\Gamma/2)^2 } \sin^2( \beta t /2
+ \phi ) e^{-\Gamma t / 2}  
\, ,
\label{eq:2res_b2t_srong}
\end{equation}
where 
\begin{equation}
\cos\phi = \frac{ \Gamma / 2 }{ \sqrt{2} \Omega } .  \label{eq:cosphi}
\end{equation}
What is interesting here are the oscillations which are given by the square
of a sine function, i.e.\ the probability oscillates as the fourth power of
a sine function which is damped at the rate $\Gamma$. In the limit $\Omega
\gg \Gamma$ the angle $\phi$ approaches $\pi/2$ and Eq.~(\ref%
{eq:2res_b2t_srong}) reduces to $b_{2}(t) \approx \cos^2( \Omega t /\sqrt{2}
) e^{-\Gamma t / 2} $.

Conversely, for weakly coupled systems, $2\Omega ^{2}<(\Gamma /2)^{2}$, the
time evolution in Eq.~(\ref{eq:2res_b2t}) can be re-expressed in the form 
\begin{equation}
b_{2}(t)=\frac{2\Omega ^{2}}{(\Gamma /2)^{2}-2\Omega ^{2}}\sinh ^{2}(\gamma
t/2+\xi )e^{-\Gamma t/2}  
\, ,
\label{eq:2res_b2t_weak}
\end{equation}%
where $\gamma ^{2}=(\Gamma /2)^{2}-2\Omega ^{2}$ and 
\begin{equation}
\cosh \xi =\frac{\Gamma /2}{\sqrt{2}\Omega }.  \label{eq:coshxi}
\end{equation}%
In the extreme limit of $\Omega \ll \Gamma $, Eq.~(\ref{eq:2res_b2t_weak})
reduces to the Fermi Golden-rule result: $b_{2}(t)\approx \exp (-2\Omega
^{2}t/\Gamma )$.

\section{Conclusion}

\label{sec:conc}

The dynamical behaviour of a three level atom in a cascade configuration in
which both transitions are coupled to a single structured reservoir of
electromagnetic field modes, and initially in the upper state, has been
analysed via Laplace transform methods. This situation involves a two photon
excitation of the reservoir, and our equations take into account the two
possible sequences in which these two photons are emitted. We have shown
that the atomic density operator is determined from the solutions of
integral equations, in which the properties of the structured reservoir only
appear via so-called reservoir structure functions, all essentially given by
the product of the mode density times the square of coupling constants. In
the cascade system two distinct reservoir structure functions are involved
since there are two transitions. The dependence of the dynamics solely on
reservoir structure functions is the necessary condition for treating
structured reservoir problems via pseudo-mode theory, so our results suggest
that it may be possible to extend pseudo-mode theory to problems involving
more than a single photon excitation of the reservoir.

This result also shows that any existing system could be replaced by an
equivalent system, provided that the reservoir structure functions were the
same in both models. This is the basis of the treatment of superradiance in
a photonic band gap continuum \cite{Bay98b} and the general treatment of
multiphoton excitation in terms of quasimodes given in our earlier work \cite%
{Dalton01a}.

In addition, a similar treatment of the dynamical behaviour of a three level
atom in a cascade configuration coupled to two separate structured
reservoirs of electromagnetic field modes, and initially in the upper state,
has been carried out. One reservoir is coupled to the upper transition, the
other to the lower transition. This situation again involves a two photon
excitation of the reservoir, but now only one possible photon emission
sequence is involved. In this situation, the equations are simpler, and the
integral equation for the amplitudes can be solved analytically. Again, the
dynamical features only depend on reservoir structure functions.

A numerical method of solving the integral equations based on discretising
the frequency space has also been obtained, and which can be applied to
various structured reservoir situations - such as for high Q cavities and
photonic band gap systems. Here we have applied this method in a numerical
test for a high Q cavity situation, where the same Lorentzian reservoir
structure function applies for both transitions, showing the non-Markovian
decay of the excited state. Results for both the single structured reservoir
case and the two separate reservoirs case have been obtained, showing the
different behaviour in the two cases. This difference is to be expected, as
the two photons emitted should have similar frequencies, and only in the two
separate reservoirs cases should it be possible to distinguish which order
the photons were emitted. In this latter case we were able to solve the
model problem analytically. Finally, a formal solution of the integral
equations based on the biorthogonal left and right eigenfunctions of the
non-Hermitean kernel has been presented for completeness in the Appendices.

Our treatment of the cascade system coupled to a structured reservoir may be
compared to those of Ref.~\cite{Bay98c} in the two separate reservoirs case
and to Ref.~\cite{Nikolopoulos00} in the single reservoir case. Both these
papers also demonstrate non-Markovian decay of the excited state. Our
fundamental amplitude equations in sections \ref{sec:analysis} and \ref{sec:
separate res} agree with those of these authors. The work in Ref.~\cite%
{Nikolopoulos00} differs from our treatment, being based on replacing the
structured reservoir with discrete modes and then using numerical methods.
The work in Ref.~\cite{Bay98c} is analytic. However, a direct comparison of
the numerical results is not yet possible with either Ref.~\cite%
{Nikolopoulos00} or Ref.~\cite{Bay98c}, since both applied their theory to a
photonic band gap system whereas our present application is for the equally
important situation of a high Q cavity. Further applications of our theory
involving good analytic approximations to the reservoir structure functions
for photonic band gap systems will, however, enable more detailed
comparisons to be made.

\section*{Acknowledgements}

We would like to acknowledge funding from the UK Engineering and Physical
Sciences Research Council and helpful discussions with J.D. Cresser.

\appendix

\section{Integral equation kernel and its eigenfunctions}

\label{sec:appA}

The kernel $K(\omega _{\lambda },\omega _{\mu })$\ involved in the integral
equation (\ref{c33.2}) and given by Eq. (\ref{c33.3}) may now be used to
define an integral operator $\hat{K}.$ The effect of $\hat{K}$\ on any
function $\phi $ is defined by 
\begin{equation}
(\hat{K}\phi )_{\omega _{\lambda }}=\int d\omega _{\mu }K(\omega _{\lambda
},\omega _{\mu })\phi (\omega _{\mu })\;.  \label{c33.4}
\end{equation}

The eigenfunctions $\phi _{n}(\omega _{\lambda })$ and eigenvalues $\xi _{n}$
for the integral operator $\hat{K}$ then satisfy 
\begin{equation}
\hat{K}\phi _{n}=\xi _{n}\phi _{n},  \label{c33.5}
\end{equation}%
or (in full) 
\begin{equation}
\int d\omega _{\mu }K(\omega _{\lambda },\omega _{\mu })\phi _{n}(\omega
_{\mu })=\xi _{n}\phi _{n}(\omega _{\lambda })\;.  \label{c34.1}
\end{equation}%
Note that we are following Ref.\ \cite{Siegman89} in our definition of the
eigenvalue of the integral equation, rather than the definition used in many
mathematical textbooks (e.g.\ Ref.\ \cite{Kanwal71}) where $1/\xi _{n}$
would be the equivalent eigenvalue.

Similarly to Eq. (\ref{c33.4}) we can define the adjoint operator $\hat{K}%
^{\dagger }$ via 
\begin{equation}
(\hat{K}^{\dagger }\phi )_{\omega _{\lambda }}=\int d\omega _{\mu }K^{\ast
}(\omega _{\mu },\omega _{\lambda })\phi (\omega _{\mu }),  \label{c34.2}
\end{equation}%
with eigenfunctions $\theta _{n}(\omega _{\lambda })$ so that 
\begin{equation}
\hat{K}^{\dagger }\theta _{n}=\xi _{n}^{\ast }\theta _{n}\;.  \label{c34.3}
\end{equation}%
It is straightforward to show that $\hat{K}^{\dagger }$ has eigenvalues
which are complex conjugates of those for $\hat{K}$ (see Appendix \ref%
{sec:appB} for details). As $\hat{K}$ will in general be non-Hermitian, the
eigenfunctions $\phi _{n}$ do not satisfy standard orthogonality conditions.
Instead the $\phi _{n}$ and the $\theta _{n}$ satisfy so-called
biorthogonality conditions 
\begin{equation}
\int d\omega _{\mu }\theta _{n}^{\ast }(\omega _{\lambda })\phi _{m}(\omega
_{\lambda })=\delta _{nm}\;.  \label{eq:biorthoganality}
\end{equation}%
The normalization result of unity for $n=m$ can be arranged by scaling
either the $\theta _{n}$ or $\phi _{m}$ by appropriate factors. Although
these results are familiar in regard to the mode functions for unstable
optical systems (\cite{Siegman89,Brown02}), they are not widely used in
quantum optics. So, for completeness, a derivation of Eq. (\ref%
{eq:biorthoganality}) is presented in Appendix \ref{sec:appC}. A formal
method of determining the eigenfunctions $\phi _{n}$ and $\theta _{n}$ is
set out in Appendix \ref{sec:appB}.


\section{Representation of the kernel}

\label{sec:appB}

We expand $\phi _{n}$ in an orthonormal basis $u_{n}$ so that 
\begin{equation}
\phi _{n}(\omega _{\lambda })=\sum_{m}\alpha _{m}^{n}u_{m}(\omega _{\lambda
})  \label{c34.4}
\end{equation}%
with 
\begin{equation}
\int d\omega _{\lambda }u_{l}^{\ast }(\omega _{\lambda })u_{m}(\omega
_{\lambda })=\delta _{lm}\;.  \label{c34.5}
\end{equation}%
Then we can write Eq. (\ref{c34.1}) as 
\begin{equation}
\sum_{m}\int d\omega _{\mu }\alpha _{m}^{n}K(\omega _{\lambda },\omega _{\mu
})u_{m}(\omega _{\mu })=\xi _{n}\sum_{m}\alpha _{m}^{n}u_{m}(\omega
_{\lambda })\;.  \label{c34.6}
\end{equation}%
Then if we multiply by $u_{l}^{\ast }(\omega _{\lambda })$ and integrate we
find 
\begin{equation}
\sum_{m}\left( \int \int d\omega _{\lambda }d\omega _{\mu }u_{l}^{\ast
}(\omega _{\lambda })K(\omega _{\lambda },\omega _{\mu })u_{m}(\omega _{\mu
})-\delta _{lm}\xi _{n}\right) \alpha _{m}^{n}=0,  \label{c35.1}
\end{equation}%
which must be true for all values of $l$. Equation (\ref{c35.1}) is a matrix
eigenvalue equation with the matrix 
\begin{equation}
K_{lm}=\int \int d\omega _{\lambda }d\omega _{\mu }u_{l}^{\ast }(\omega
_{\lambda })K(\omega _{\lambda },\omega _{\mu })u_{m}(\omega _{\mu })
\label{c35.2}
\end{equation}%
and eigenvalues which satisfy 
\begin{equation}
\left\vert K_{lm}-\xi \delta _{lm}\right\vert =0\;.  \label{c35.3}
\end{equation}%
For the operator $\hat{K}^{\dagger }$ (see Eq. (\ref{c34.2})) the matrix is
replaced by its adjoint and clearly its eigenvalues are complex conjugates
of those for $\hat{K}$.

The explicit form for $K_{lm}$ is found by substitution of the expressions (%
\ref{c33.3}) and (\ref{c32.1.3}) into Eq. (\ref{c35.2}) which yields 
\begin{eqnarray}
K_{lm} &=&\int \int d\omega _{\lambda }d\omega _{\mu }u_{l}^{\ast }(\omega
_{\lambda })\Biggl[  \nonumber \\
&&\times \frac{1}{A(\omega _{\lambda })}\left( \rho (\omega _{\mu })g_{\mu
1}^{2}\frac{g_{\lambda 1}g_{\mu 2}}{g_{\lambda 2}g_{\mu 1}}\frac{1}{%
s+i(\omega _{\lambda }+\omega _{\mu }-\omega _{1}-\omega _{2})}+\frac{\rho
(\omega _{\mu })g_{\mu 2}^{2}}{s}\right) \Biggr]u_{m}(\omega _{\mu })\;.
\label{c35.4}
\end{eqnarray}%
The integral over $\omega _{\mu }$ will involve the reservoir structure
functions as defined in Eq. (\ref{eq.res.struc.fns}). The function $A(\omega
_{\lambda })$ is also obtainable from the reservoir structure functions (see
Eq. (\ref{c32.1.2})). Note that $g_{\lambda 1}g_{\mu 2}/g_{\lambda 2}g_{\mu
1}$ is independent of frequency in Eq. (\ref{c35.4}).

In summary, the matrix $K_{lm}$ and hence the eigenfunctions $\phi
_{n},\theta _{n}$ and eigenvalues are all obtained from the reservoir
structure functions and given functions, such as the basis set $u_{n}$.


\section{Bi-orthogonality of eigenfunctions}

\label{sec:appC}

To show that the eigenfunctions satisfy a biorthogonality condition we first
write from Eqs. (\ref{c33.5}) and (\ref{c34.1}, \ref{c34.2}, \ref{c34.3}) 
\begin{eqnarray}
\int d\omega _{\mu }K(\omega _{\lambda },\omega _{\mu })\phi _{n}(\omega
_{\mu }) &=&\xi _{n}\phi _{n}(\omega _{\lambda })  \nonumber \\
\int d\omega _{\mu }K(\omega _{\mu },\omega _{\lambda })\theta _{m}^{\ast
}(\omega _{\mu }) &=&\xi _{m}\theta _{m}^{\ast }(\omega _{\lambda })\;.
\label{c36.1}
\end{eqnarray}%
After multiplying the first equation by $\theta _{m}^{\ast }(\omega
_{\lambda })$, the second by $\phi _{n}(\omega _{\lambda })$ and then
integrating over $\omega _{\lambda }$ we find that 
\begin{eqnarray}
\int \int d\omega _{\lambda }d\omega _{\mu }\theta _{m}^{\ast }(\omega
_{\lambda })K(\omega _{\lambda },\omega _{\mu })\phi _{n}(\omega _{\mu })
&=&\xi _{n}\int d\omega _{\lambda }\theta _{m}^{\ast }(\omega _{\lambda
})\phi _{n}(\omega _{\lambda })  \nonumber \\
\int \int d\omega _{\lambda }d\omega _{\mu }\phi _{n}(\omega _{\lambda
})K(\omega _{\lambda },\omega _{\mu })\theta _{m}^{\ast }(\omega _{\mu })
&=&\xi _{m}\int d\omega _{\lambda }\theta _{m}^{\ast }(\omega _{\lambda
})\phi _{n}(\omega _{\lambda })\;.  \label{c36.2}
\end{eqnarray}%
After a change of variable in the second equation, the left-hand sides are
equal and we then conclude that 
\begin{equation}
(\xi _{n}-\xi _{m})\int d\omega _{\lambda }\theta _{m}^{\ast }(\omega
_{\lambda })\phi _{n}(\omega _{\lambda })=0,  \label{c36.3}
\end{equation}%
so that the biorthogonality condition 
\begin{equation}
\int d\omega _{\lambda }\theta _{m}^{\ast }(\omega _{\lambda })\phi
_{n}(\omega _{\lambda })=0  \label{c36.4}
\end{equation}%
applies unless $\xi _{m}=\xi _{n}$.


\section{Integral equation solution in terms of eigenfunctions of $K$}

\label{sec:appD}

We will \textit{assume} that the set of eigenfunctions $\phi _{n}$ form a
basis for expanding the solution $\overline{f}(\omega _{\lambda })$ (to Eq. (%
\ref{c33.2})). Likewise we will assume that $d(\omega _{\lambda })$ can be
expanded in terms of the $\phi _{n}$ so that 
\begin{eqnarray}
\overline{f}(\omega _{\lambda }) &=&\sum_{n}\overline{f}_{n}\phi _{n}(\omega
_{\lambda })  \nonumber \\
d(\omega _{\lambda }) &=&\sum_{n}d_{n}\phi _{n}(\omega _{\lambda }).
\label{c37.2}
\end{eqnarray}%
Using the biorthogonality of the eigenfunctions (Eq. (\ref%
{eq:biorthoganality})) the expansion coefficients can be found as: 
\begin{eqnarray}
\overline{f}_{n} &=&\int d\omega _{\lambda }\theta _{n}^{\ast }(\omega
_{\lambda })\overline{f}(\omega _{\lambda })  \nonumber \\
d_{n} &=&\int d\omega _{\lambda }\theta _{n}^{\ast }(\omega _{\lambda
})d(\omega _{\lambda }).  \label{c37.3}
\end{eqnarray}%
Substituting from Eq. (\ref{c37.2}) into Eq. (\ref{c33.2}) and using the
eigenvalue equation (\ref{c34.1}) we find that: 
\begin{eqnarray}
\sum_{n}\overline{f}_{n}\phi _{n}(\omega _{\lambda })+\sum_{n}\overline{f}%
_{n}\int d\omega _{\mu }K(\omega _{\lambda },\omega _{\mu })\phi _{n}(\omega
_{\mu }) &=&\sum_{n}d_{n}\phi _{n}(\omega _{\lambda })  \nonumber \\
\sum_{n}(\overline{f}_{n}+\xi _{n}\overline{f}_{n}-d_{n})\phi _{n}(\omega
_{\lambda }) &=&0.  \label{c37.4}
\end{eqnarray}%
Using the biorthogonality result for the eigenvalue $\xi _{n}$ we see that 
\begin{equation}
\overline{f}_{n}(1+\xi _{n})-d_{n}=0,  \label{c38.1}
\end{equation}%
so that provided $\xi _{n}\neq -1$ 
\begin{equation}
\overline{f}_{n}=\frac{d_{n}}{1+\xi _{n}},  \label{c38.2}
\end{equation}%
which gives the solutions for the expansion coefficients for $\overline{f}%
(\omega _{\lambda })$ in terms of known quantities. The quantities $%
\overline{f}_{n},\phi _{n}(\omega _{\lambda }),K(\omega _{\lambda },\omega
_{\mu })$ and $\xi _{n}$ are of course all functions of the Laplace variable 
$s$, but for simplicity of notation $s$ is left implicit.

\pagebreak

\section*{Figures}

\end{document}